\documentclass{article}
\usepackage{spconf,amsmath,graphicx,hyperref}

 \usepackage{amsthm}
\usepackage{amssymb}
 \usepackage{algorithm}
 \usepackage{algorithmic}
 \usepackage{acronym}
\usepackage{subfigure}

\acrodef{adc}[ADC]{analog-to-digital convertor}
\acrodef{ris}[RIS]{reconfigurable intelligent surface}
\acrodef{bs}[BS]{base station}
\acrodef{ue}[UE]{User Equipment}
\acrodef{isac}[ISAC]{integrated sensing and communication}
\acrodef{upa}[UPA]{uniform planar array}

\DeclareMathOperator{\diag}{diag}

\def\bd{{\mathbf d}}
\def\bD{{\mathbf D}}

\def\p{{\mathbf p}}
\def\bT{{\mathbf T}}
\def\bX{{\mathbf X}}
\def\bQ{{\mathbf Q}}
\def\bR{{\mathbf R}}
\def\bu{{\mathbf u}}
\def\bU{{\mathbf U}}

\def\bB{{\mathbf B}}
\def\bS{{\mathbf S}}
\def\H{{\text H}}
\def\V{{\text V}}
\def\brho{{\boldsymbol{\rho}}}
\title{A TWO-DIMENSIONAL SUPER-RESOLUTION METHOD FOR RECONFIGURABLE INTELLIGENT SURFACE-ASSISTED NEAR-FIELD LOCALIZATION}
\twoauthors
 {Feng Xi}
	{Department of Electronic Engineering\\
	Nanjing University of Science and Technology\\
	Nanjing 210094, China}
 {Dehui Yang}
	{School of Mathematics and Physics\\
	Xi’an Jiaotong-Liverpool University\\
	Suzhou 215123, China}
\begin{document}
%
\maketitle
\begin{abstract}
Reconfigurable intelligent surface (RIS)–aided localization in the radiating near‑field requires range‑aware spherical-wave models, which inherently couple angles and ranges and thus complicate accurate 3D positioning. Using the Fresnel approximation, we show that the RIS response can be expressed as the element-wise product of a 2D far‑field steering vector and a range‑dependent quadratic‑phase chirp. By modeling these chirp components within a low‑dimensional subspace, we reformulate the joint recovery of azimuth, elevation, and range under a 2D super‑resolution framework, resulting in a 2D atomic norm minimization (2D-ANM) problem. Solving this via semi-definite programming (SDP) yields gridless azimuth-elevation estimation and high-accuracy range recovery. Simulations demonstrate accurate 3D localization and enhanced robustness of the proposed scheme, compared with subspace and compressive sensing methods.
\end{abstract}


\section{Introduction}
\label{sec:intro}
Reconfigurable intelligent surfaces (RISs) are low-power, nearly passive metasurfaces that program element-wise phases to reshape radio propagation, enabling robust connectivity and high-accuracy localization even under non-line-of-sight conditions \cite{Renzo-JSAC20,Emil-SPM22}.
As apertures scale and carriers move to mmWave/THz bands, users increasingly operate within the radiating near-field (NF) region of large RISs, where wavefronts are spherical and range-dependent.
In this regime, conventional far-field (FF) plane-wave models are no longer accurate.
While the spherical-wave model supports 3D positioning with a uniform planar array (UPA) RIS \cite{Wymeersch-VTM20,Pan-JSTSP23,Yan-TGCN25}, joint estimation of azimuth, elevation, and range is challenging due to the high dimensionality of the parameter space and strong nonlinear coupling among these parameters.

Radio localization is central to array processing and a key enabling technology for \ac{isac} in 6G \cite{Liu-JSAC22}.
Two representative approaches are subspace \cite{stoica2005spectral} and compressive sensing (CS) methods \cite{Malioutov-TSP05}. 
Subspace approaches rely on accurate sample-covariance estimation and typically require many snapshots alongside multi-dimensional search. CS approaches discretize continuous parameters to obtain a sparse linear model, but suffer from grid mismatch and resolution loss.
In RIS-assisted localization, integrating a very large number of low-cost elements on a single panel places user equipment (UE) frequently in the RIS's NF region, necessitating a shift from the FF planar-wave to the NF spherical-wave model.
Meanwhile, RISs provide only reflective control with limited on-board processing, further complicating accurate NF localization.

To address these issues, both subspace and CS methods have been tailored to the RIS-aided scenarios.
For subspace methods, \cite{Pan-JSTSP23} constructs a down-sampled Toeplitz covariance to partially decouple parameters, and
\cite{Ramezani-GLOBECOM24} exploits array symmetry with spatial smoothing to obtain a well-conditioned covariance matrix.
For CS methods, \cite{Cui-ToC22, Wei-CCom22, Liu-ICC04} design NF codebooks to facilitate orthogonal matching pursuit (OMP)-based sparse recovery, while \cite{Yan-TGCN25} combines OMP with maximum likelihood (ML) refinement to improve accuracy.
Despite these advances, achieving high-resolution multi-dimensional estimation from few snapshots remains difficult due to the strong parameter coupling.

This paper develops a super-resolution framework for RIS-assisted NF localization by exploiting the Fresnel approximation and the inherent structure of the UE-RIS channel.
Under the Fresnel model, the RIS response can be expressed as a 2D FF steering vector modulated by a range-dependent quadratic-phase chirp.
For a UPA RIS, the response is further separable across the horizontal and vertical axes via a Kronecker structure.
Leveraging these properties and modeling the chirps with low-dimensional subspaces, we reformulate the RIS‑assisted NF localization as a 2D super‑resolution problem \cite{Yang-TIT16} that simultaneously recovers the 2D FF steering vectors and the unknown chirp components. 
As a result, this problem can be solved via 2D atomic norm minimization (2D-ANM), yielding gridless azimuth–elevation estimation and high-accuracy range recovery.
Our approach generalizes the 2D super-resolution framework \cite{Chi-TSP2015,Suliman-TSP22} from FF to NF channels and extends our prior work on 1D NF estimation \cite{Li-ICASSP25}.
Simulations show that the proposed method achieves superior performance over subspace and CS methods, especially in regimes with limited snapshots.

\section{Signal Model}
\label{sec:format}
We consider a RIS-aided \ac{isac} system in which a \ac{bs} with $M$ antennas communicates via an $N$-element RIS.
The RIS forms a UPA with $N_\H$ elements per row and $N_\V$ elements per column ($N=N_\H N_\V$).
Both $N_\H$  and $N_\V$ are assumed to be odd integers, and the element at the geometric center serves as the reference.
The RIS lies on the $YZ$-plane with its array axes aligned to the $y$- and $z$-directions. 
Let $d_\H$ and $d_\V$ be the inter-element spacings along the $y$- and $z$-axes, respectively.
Index elements by ($n_\H$, $n_\V$) with $n_\H\in\mathcal{N}_\H = \{-\frac{N_\H-1}{2},\cdots,\frac{N_\H-1}{2}\}$, $n_\V\in\mathcal{N}_\V = \{-\frac{N_\V-1}{2},\cdots,\frac{N_\V-1}{2}\}$, and map ($n_\H$,$n_\V$) to the scalar index $n =(n_\H+\frac{N_\H-1}{2})N_\V + (n_\V + \frac{N_\V-1}{2})+1$.
For notational simplicity, we use $n$ and $(n_\H,n_\V)$ interchangeably hereafter.

The system aims to localize $K$ single-antenna UEs operating in the radiating NF of the RIS. We consider the scenario where direct UE-\ac{bs} links are blocked, so both uplink and downlink occur via the RIS.
The spherical coordinate of the $k$th UE is $(r_k,\psi_k,\theta_k)$, i.e., its Cartesian coordinate $\p_k = r_k(\cos(\psi_k)\cos(\theta_k),\sin(\psi_k)\cos(\theta_k),\sin(\theta_k))$,
where $r_k$ is the distance from the $k$th UE to the reference RIS element, and $(\psi_k,\theta_k)$ are the azimuth and elevation angles of arrival (AOAs) at the RIS.

Denote the RIS-BS channel as $\mathbf{H}\in \mathbb{C}^{M\times N}$, which is assumed to be static and known since the \ac{bs} and the RIS are stationary.
Then the received signal at the BS during the $t$th time-slot is
\begin{equation}\label{eqn:yt}
    \mathbf{y}(t)=\sum_{k=1}^K \sqrt{p_k} \mathbf{H}\boldsymbol{\Phi}(t)\mathbf{g}_ks_k(t) + \mathbf{w}(t), 
\end{equation}
where $p_k$ is the transmit power of the $k$th UE, $s_k(t)$ is the pilot symbol, $\mathbf{w}(t)$ is the additive random noise, and $\boldsymbol{\Phi}(t)= \mathrm{diag}(e^{j\phi_1(t)}, \cdots,e^{j\phi_N(t)} )$ is the RIS reflection matrix with $\phi_n(t)$ being the phase shift applied by the $n$th RIS element at the time-slot $t$.
The UE-RIS channel is $\mathbf{g}_k = \eta_k \mathbf{b}(\psi_k, \theta_k, r_k)\in\mathbb{C}^N$, with path loss $\eta_k$ and RIS response $\mathbf{b}(\psi_k, \theta_k, r_k)$.
Let $r_k^{(n)}$ denote the distance from the $k$th UE to the $n$th RIS element.
The $n$th element of $\mathbf{b}(\psi_k, \theta_k, r_k)$ is given by
\begin{equation}\label{eqn: bn}
    [\mathbf{b}(\psi_k, \theta_k, r_k)]_n = \exp\left\{j\frac{2\pi}{\lambda}\left(r_k - r_k^{(n)}\right)\right\},
\end{equation}
where $\lambda$ is the carrier wavelength.

In the radiating NF region, $r_k^{(n)}$ admits a Fresnel approximation: 
\begin{equation}
\begin{split}
    r_k^{(n)} \approx & r_k - n_\H d_\H\sin(\psi_k)\cos(\theta_k) - n_\V d_\V\sin(\theta_k)\\& + \frac{n_\H^2d_\H^2+ n_\V^2d_\V^2}{2r_k}.
\end{split}
\end{equation}
Consequently, (\ref{eqn: bn}) can be approximated as
\begin{equation}
\begin{split}
&\left[\mathbf{b}(\psi_k, \theta_k, r_k)\right]_{n}\\ \approx &\exp\left\{j\left((n_\H \alpha_k+n_\V\beta_k) - (n_\H^2d_\H^2 + n_\V^2d_\V^2)\gamma_k\right)\right\},    
\end{split}
\end{equation}
with spatial frequencies $\alpha_k = \frac{2\pi}{\lambda}d_\H\sin(\psi_k)\cos(\theta_k)$, $\beta_k =\frac{2\pi}{\lambda}d_\V\sin(\theta_k)$, and $\gamma_k = \frac{\pi}{\lambda r_k}$. 

Define the far‑field steering vectors $\mathbf{a}_\H(\alpha)$, $\mathbf{a}_\V(\beta)$ and the quadratic‑phase chirp vectors $\mathbf{q}_\H(\delta)$, $\mathbf{q}_\V(\delta)$ as
\begin{align*}
    \mathbf{a}_\H(\alpha)=\big[e^{j n_\H \alpha}\big]_{n_\H\in\mathcal{N}_\H},\quad &\mathbf{a}_\V(\beta)=\big[e^{j n_\V \beta}\big]_{n_\V\in\mathcal{N}_\V},\\
    \mathbf{q}_\H(\delta)=\big[e^{-j n_\H^2 \delta}\big]_{n_\H\in\mathcal{N}_\H},\quad &\mathbf{q}_\V(\delta)=\big[e^{-j n_\V^2 \delta}\big]_{n_\V\in\mathcal{N}_\V}.
\end{align*}
The RIS response can be compactly rewritten as
\begin{equation}\label{eqn:b}
\begin{split}
     &\mathbf{b}(\psi_k, \theta_k, r_k)\\\approx& (\mathbf{a}_\H(\alpha_k)\otimes \mathbf{a}_\V(\beta_k)) \odot (\mathbf{q}_\H(d_\H^2 \gamma_k)\otimes \mathbf{q}_\V(d_\V^2 \gamma_k))\\
     = &(\mathbf{a}_\H(\alpha_k)\odot \mathbf{q}_\H(d_\H^2 \gamma_k)) \otimes (\mathbf{a}_\V(\beta_k)\odot \mathbf{q}_\V(d_\V^2 \gamma_k)),
\end{split}
\end{equation}
where $\otimes$ and $\odot$ denote the Kronecker and Hadamard products, respectively. 
Thus, estimating the original parameters $\{\psi_k, \theta_k, r_k\}_{k=1}^K$ is equivalent to recovering the parameters $\{\alpha_k, \beta_k, \gamma_k\}_{k=1}^K$.
Expression (\ref{eqn:b}) explicitly separates the linear (angle-dependent) and quadratic (range-dependent) phases along the horizontal and vertical axes, which will be exploited for joint azimuth-elevation-range estimation.


\section{Two-Dimensional Super-Resolution Framework}
In this section, we exploit the separable structure in (\ref{eqn:b}) to reformulate the RIS-assisted NF localization problem under a 2D super-resolution framework and solve it via 2D-ANM.

Following \cite{Li-ICASSP25},
we approximate the quadratic-phase terms by low-dimensional subspaces, i.e., 
\begin{align}\label{eqn:subspace}
    \mathbf{q}_\H(d_\H^2 \gamma_k) \approx \mathbf{B}_\H \mathbf{u}_\H^k, \quad \mathbf{q}_\V(d_\V^2 \gamma_k) \approx \mathbf{B}_\V \mathbf{u}_\V^k,
\end{align}
where $\mathbf{B}_\H\in\mathbb{C}^{N_\H \times J_\H}$, $\mathbf{B}_\V\in\mathbb{C}^{N_\V \times J_\V}$ are subspace matrices with $J_\H \ll N_\H$ and $J_\V \ll N_\V$, and $\mathbf{u}_\H^k$ and $\mathbf{u}_\V^k$ are the corresponding coefficients.
The subspaces can be constructed, for example, by discretizing the quadratic phase \cite{Li-ICASSP25} or using discrete prolate spheroidal sequences (DPSSs) \cite{Slepian-1978,Zhu-2017}.

Let $(\mathbf{b}_\H^{(n_\H)})^H$, $(\mathbf{b}_\V^{(n_\V)})^H$ be the $n_\H$th, $n_\V$th rows of $\mathbf{B}_\H$ and $\mathbf{B}_\V$, respectively, and $\mathbf{e}_\H^{(n_\H)}$, $\mathbf{e}_\V^{(n_\V)}$ be the corresponding canonical basis vectors.
Define the linear operator $\mathcal{P}(\cdot): \mathbb{C}^{J_\H J_\V\times N} \mapsto \mathbb{C}^N$ with its $n$th element given by
\begin{equation}
    [\mathcal{P}(\mathbf{Z})]_n = \langle \mathbf{Z}, \mathbf{b}_\H^{(n_\H)}(\mathbf{e}_\H^{(n_\H)})^H \otimes \mathbf{b}_\V^{(n_\V)}(\mathbf{e}_\V^{(n_\V)})^H \rangle,
\end{equation}
where $\langle \mathbf{A}, \mathbf{B}  \rangle = \text{Tr}(\mathbf{B}^H\mathbf{A})$.
Using $\langle \mathbf{A}\otimes \mathbf{B} , \mathbf{C}\otimes \mathbf{D} \rangle = \langle \mathbf{A},\mathbf{C} \rangle \langle \mathbf{B},\mathbf{D} \rangle$, the RIS response (\ref{eqn:b}) becomes
\begin{equation}
    \mathbf{b}(\psi_k, \theta_k, r_k) = \mathcal{P}\left(\mathbf{u}_\H^k\mathbf{a}_\H^T(\alpha_k) \otimes \mathbf{u}_\V^k\mathbf{a}_\V^T(\beta_k)\right).
\end{equation}

Define the rank-$K$ matrix
\begin{equation}\label{eqn:X_o}
\begin{split}
     \mathbf{X}_o &= \sum_{k=1}^K \tilde{\eta}_k \mathbf{u}_\H^k\mathbf{a}_\H^T(\alpha_k) \otimes \mathbf{u}_\V^k\mathbf{a}_\V^T(\beta_k) \\&= \sum_{k=1}^K \tilde{\eta}_k (\mathbf{u}_\H^k \otimes \mathbf{u}_\V^k) (\mathbf{a}_\H(\alpha_k) \otimes \mathbf{a}_\V(\beta_k))^T,
\end{split}
\end{equation}
with $\tilde{\eta}_k = \sqrt{p_k} \eta_k$.
Without loss of generality, set the pilot symbols $s_k(t)=1$ for $t=1,\cdots,L$.
Stacking 
$\mathbf{y} = [\mathbf{y}^T(1), \cdots, \mathbf{y}^T(L)]^T$ yields
\begin{equation}\label{eqn:y_proj}
\begin{split}
        \mathbf{y} = \bar{\mathbf{H}} \mathcal{P}(\mathbf{X}_o) + \mathbf{w},
\end{split}
\end{equation}
where $\bar{\mathbf{H}} = [(\mathbf{H}\boldsymbol{\Phi}(1))^T, \cdots,(\mathbf{H}\boldsymbol{\Phi}(L))^T]^T$ and $\mathbf{w} = [\mathbf{w}^T(1),\\ \cdots, \mathbf{w}^T(L)]^T$.
Therefore, the parameter estimation problem reduces to recovering the low-rank matrix $\mathbf{X}_o$ from $\mathbf{y} = \bar{\mathbf{H}} \mathcal{P}(\mathbf{X}_o)$.
Note that $\mathbf{X}_o$ is a multiple measurement vector (MMV) model with 2D line spectra $\{(\alpha_k, \beta_k)\}_{k=1}^K$,
which can be efficiently solved by 2D atomic norm minimization.

Let $\mathbf{d}(\alpha,\beta) = \mathbf{a}_\H(\alpha) \otimes \mathbf{a}_\V(\beta)$ and define the atomic set $\mathcal{A} = \left\{\mathbf{u} \mathbf{d}^T(\alpha,\beta) \; \big|\; \|\mathbf{u}\|_2 =1,\; \mathbf{u}\in\mathbb{C}^{J_\H J_\V},\; \alpha,\beta\in [-\pi, \pi] \right\}$.
The associated atomic norm is
\begin{equation*}
    \|\mathbf{X}\|_{\mathcal{A}} = \inf_{\eta_l, \alpha_l,\beta_l, \mathbf{u}_l} \left\{\sum_l |\eta_l|: \mathbf{X} = \sum_l \eta_l \mathbf{u}_l \mathbf{d}^T(\alpha_l, \beta_l) \right\}.
\end{equation*}
Therefore, $\mathbf{X}_o$ is recovered via solving the following convex program:
\begin{equation}\label{prob:ANM}
        \min_{\mathbf{X}} \quad \|\mathbf{X}\|_{\mathcal{A}} 
        \quad\text{s.t.}\quad \|\mathbf{y} - \bar{\mathbf{H}} \mathcal{P}(\mathbf{X})\|_2\leq \epsilon,
\end{equation}
where $\epsilon>0$ is a predefined parameter controlling the noise level.
Problem (\ref{prob:ANM}) can be cast as semidefinite programming (SDP) by exploiting the 2-D structure induced by $\bd(\alpha_k, \beta_k)$.

Define the two-fold Toeplitz matrix $\text{Toep}(\mathbf{T}) \in \mathbb{C}^{N \times N}$ with a given $\mathbf{T}\in\mathbb{C}^{(2N_\H -1)\times (2N_\V-1)}$:
\begin{equation}
\begin{aligned}
\label{tf-toep-outer}
& \text{Toep}(\bT) = \begin{bmatrix}
\bT_0 & \bT_{-1} & \cdots & \bT_{-(N_\H-1)} \\
\bT_1 & \bT_{0} & \cdots & \bT_{-(N_\H - 2)} \\
\vdots & \vdots & \ddots & \vdots \\
\bT_{N_\H - 1} & \bT_{N_\H -2 } & \cdots & \bT_0 
\end{bmatrix}, \\
\end{aligned}
\end{equation}
where each block $\bT_{\ell}$ in $\text{Toep}(\bT)$ is an $N_\V \times N_\V$ Toeplitz matrix formed by the $\ell$th row of $\bT$
with $\bT_{\ell}[i, j] = \bT[\ell, i-j],~0\leq i, j\leq N_\V-1$.

According to \cite{Chi-TSP2015}, (\ref{prob:ANM}) admits an approximate SDP characterization:
\begin{equation}\label{prob:SDP}
    \begin{split}
        \min_{\mathbf{X},\bT, \mathbf{Q}} \quad &\frac{1}{2} \text{Tr}(\mathbf{Q}) + \frac{1}{2N}\text{Tr}(\text{Toep}(\bT)) \\
        \text{s.t.}\quad &\|\mathbf{y} - \bar{\mathbf{H}} \mathcal{P}(\mathbf{X})\|_2\leq \epsilon, \\
        & \begin{bmatrix}
            \mathbf{Q} & \mathbf{X}\\
            \mathbf{X}^H& \text{Toep}(\bT)
        \end{bmatrix} \succeq 0,
    \end{split}
\end{equation}
where $\mathbf{Q}\in\mathbb{C}^{J_\H J_\V\times J_\H J_\V}$. The SDP can be solved using the off-the-shelf CVX toolbox \cite{cvx}. 
For large-scale problems, the alternating direction method of multipliers (ADMM) \cite{Ran-ICASSP21} or coordinate descent method \cite{Li-TSP24} can be used to reduce the computational complexity.

\begin{figure*}[htb]
\centering
   \begin{subfigure}[$\psi$]
    {\includegraphics[width=0.31\textwidth]{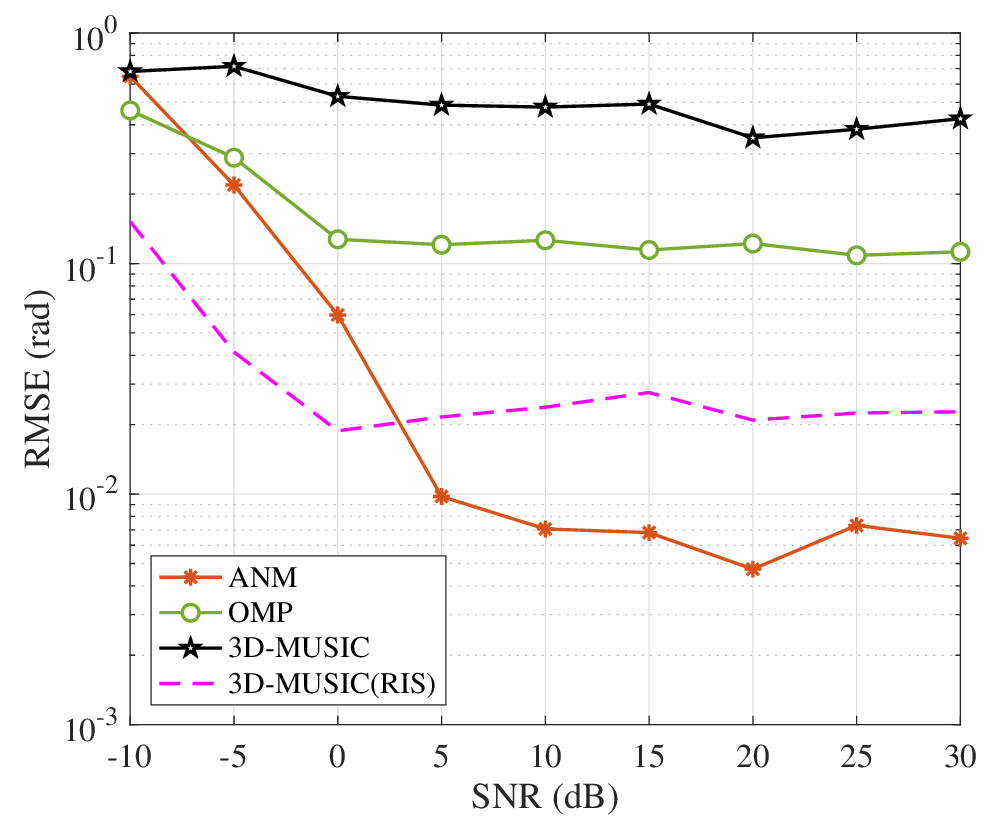}}
   \end{subfigure} 
    \begin{subfigure}[$\theta$]
    {\includegraphics[width=0.31\textwidth]{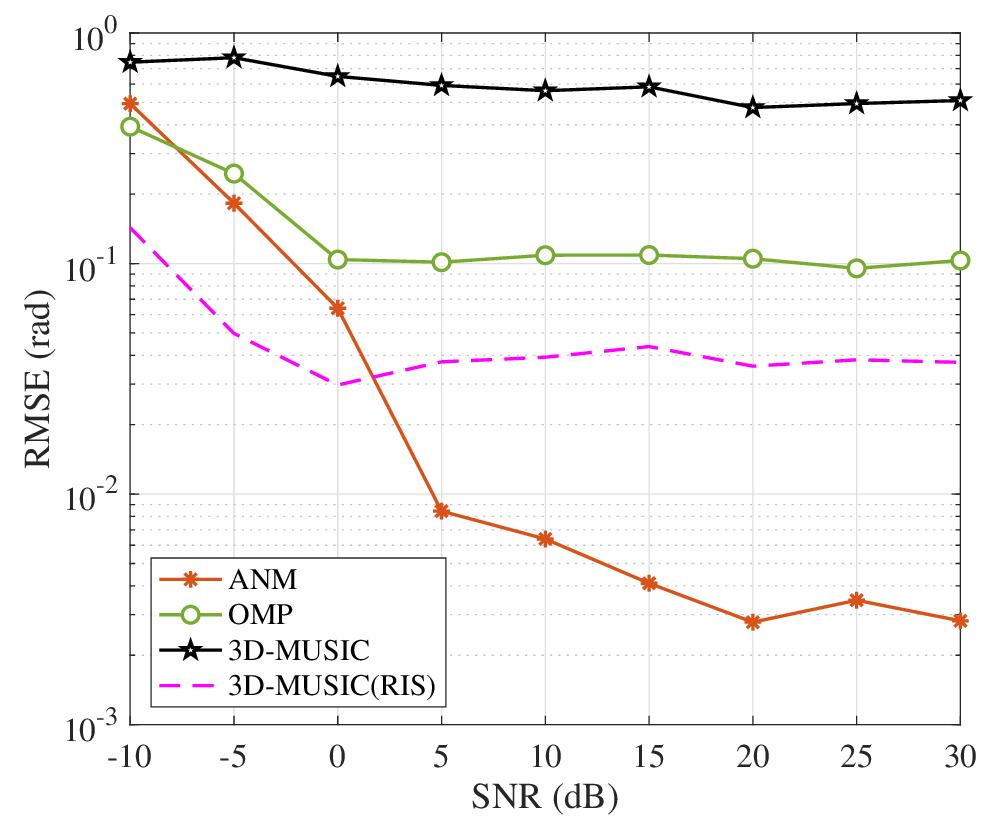}}
   \end{subfigure} 
    \begin{subfigure}[$r$]
    {\includegraphics[width=0.31\textwidth]{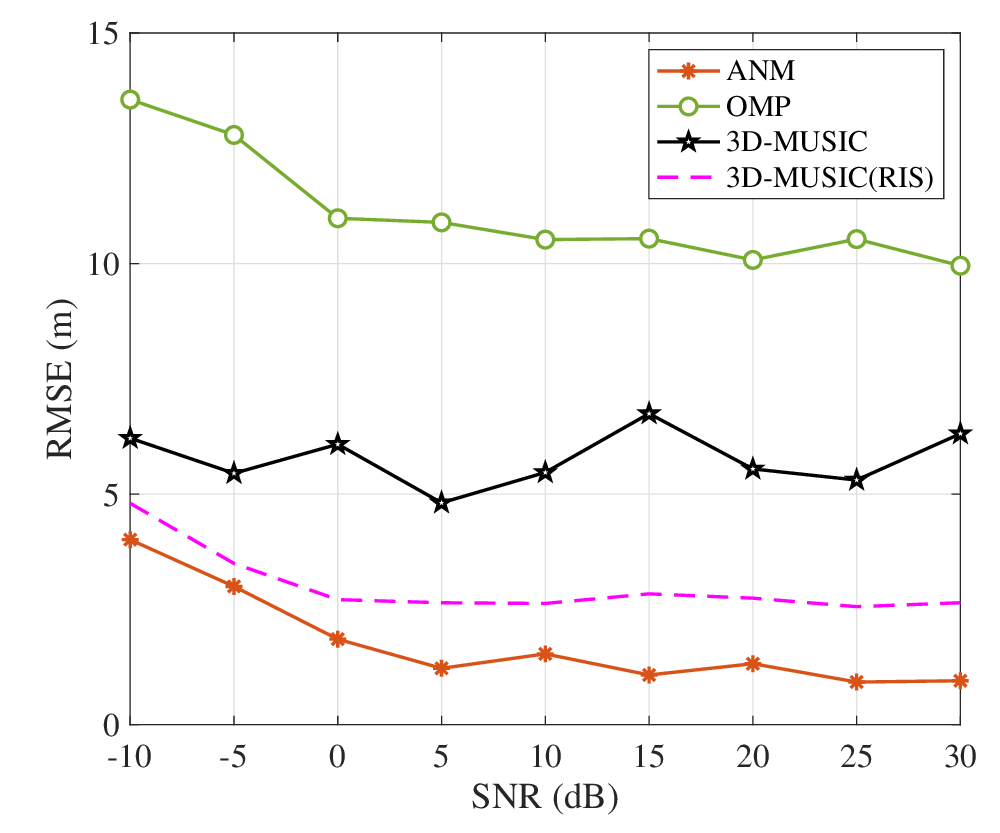}}
   \end{subfigure} 
\caption{RMSE performance of the azimuth-elevation-range estimation with respect to different SNR levels.}
\vspace{-0.3cm}
\label{fig:RMSEvsSNR}
\end{figure*}

\begin{figure}[htb]
\centering
    {\includegraphics[width=0.31\textwidth]{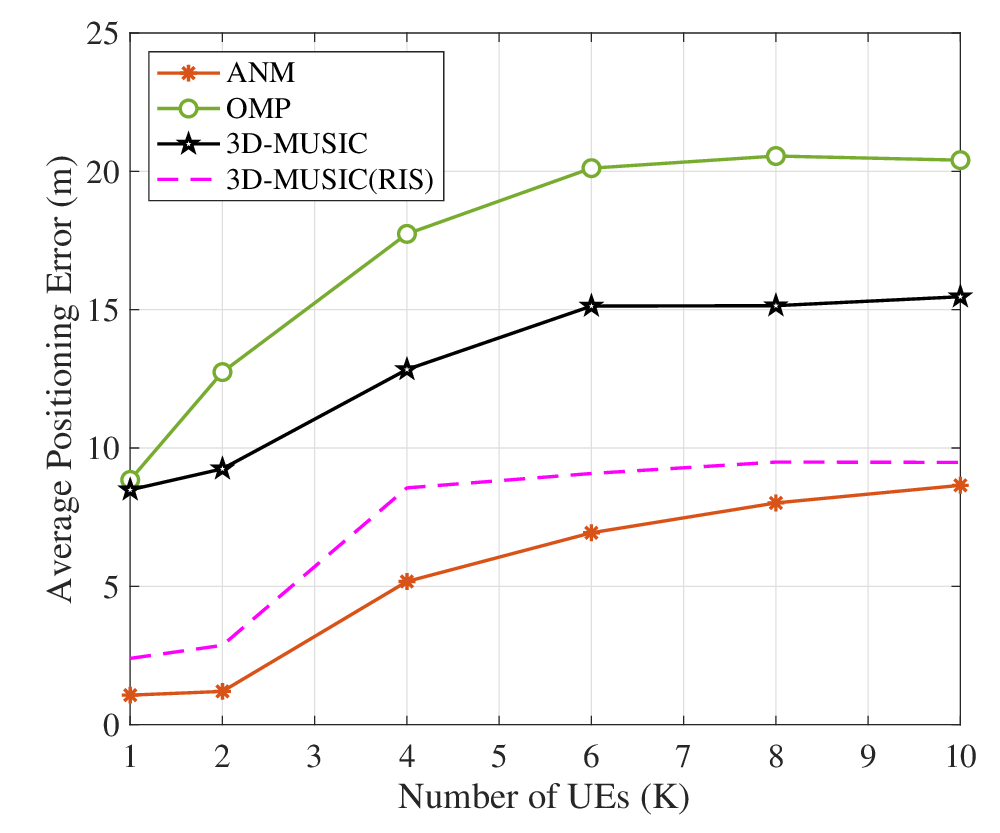}}
    \caption{Average positioning error versus the number of UEs.}
  \vspace{-0.3cm}
  \label{fig:RMSEvsK}
\end{figure} 

\section{Joint Azimuth-Elevation-Range Estimation}
\label{sec:pagestyle}
Let $(\hat{\bX},\hat{\bT},\hat{\bQ})$ denote the optimal solution to the SDP in (\ref{prob:SDP}).
We now introduce the detailed steps to recover the set of parameters $\{\eta_k,\psi_k, \theta_k, r_k\}_{k=1}^K$ from $(\hat{\bX},\hat{\bT},\hat{\bQ})$.

\noindent\textbf{1) Azimuth-elevation estimation}: 
With $\hat\bT$, we form the two-fold Toeplitz matrix
$\hat{\bR} = \text{Toep}(\hat\bT)\in \mathbb{C}^{N\times N}$. 
By the multilevel Vandermonde decomposition \cite{Yang-TIT2016}, $\hat{\bR}$ admits the following Vandermonde decomposition:
\begin{equation}
    \hat{\bR} = \bD \diag([c_1,\cdots,c_k])\bD^H, \quad c_k>0,
\end{equation}
where $\bD= [\bd(\hat{\alpha}_1,\hat{\beta}_1),\cdots,\bd(\hat{\alpha}_K,\hat{\beta}_K)]\in\mathbb{C}^{N\times K}$ is a Vandermonde matrix.
The spatial frequencies $\{(\hat{\alpha}_k,\hat{\beta}_k)\}$ are then extracted from $\hat{\mathbf{R}}$ via the matrix pencil and pairing (MaPP) procedure \cite{Yang-TIT2016}, which are then mapped to angles via $\hat{\theta}_k = \arcsin \left(\frac{\hat{\beta}_k\lambda}{2\pi d_\V} \right)$ and $\hat{\psi}_k = \arcsin \left( \frac{\hat{\alpha}_k\lambda}{2\pi d_\H \cos \hat{\theta}_k}\right)$.

\noindent\textbf{2) Range estimation}:
From \eqref{eqn:X_o}, $\mathbf{X}_o$ can be rewritten as  
$\mathbf{X}_o= \mathbf{U}\mathbf{D}^T$,
where $\mathbf{U}=[\tilde{\eta}_1(\mathbf{u}_\H^1\!\otimes\!\mathbf{u}_\V^1),\ldots,\tilde{\eta}_K(\mathbf{u}_\H^K\!\otimes\!\mathbf{u}_\V^K)]\in\mathbb{C}^{J_\H J_\V\times K}$.
Projecting $\hat{\mathbf{X}}$ onto $\mathrm{span}(\mathbf{D})$ gives the estimate
\begin{equation}
    \hat\bU = \hat\bX (\bD^T)^\dagger \in\mathbb{C}^{J_\H J_\V\times K},
\end{equation}
where $(\cdot)^\dagger$ denotes the Moore-Penrose inverse. Each column of $\hat\bU$, denoted as $\hat{\bu}_k$, approximates the separable Kronecker form $\bu_{\H}^k\otimes \bu_{\V}^k$ up to a scaling factor due to the unknown $\tilde{\eta}_k$.
Reshaping $\hat{\bu}_k$ into a $J_\H\times J_\V$ matrix and taking its best rank-one approximation $\hat{\bu}_k\approx \sigma_k \hat{\bu}_\H^k (\hat{\bu}_\V^k)^H$ yields the estimates of $\bu_{\H}^k$ and $\bu_{\V}^k$ in (\ref{eqn:subspace}).
Then we can reconstruct the quadratic-phase chirp vectors:
\begin{align}
    \hat{\mathbf{q}}_\H^k = \bB_\H\hat{\bu}_\H^k,\quad \hat{\mathbf{q}}_\V^k = \bB_\V\hat{\bu}_\V^k.    
\end{align}
Note that for a 1D chirp signal $\mathbf{q}[n] = \eta e^{-jn^2\gamma d^2}$, we have $\frac{\mathbf{q}[n+1]\mathbf{q}[n-1]}{\mathbf{q}[n]^2} = e^{-j2\gamma d^2}$.
We use this property to eliminate the effect of the scaling ambiguity. By defining
\begin{align}
    \brho_x^k [n_x] = \frac{\hat{\mathbf{q}}_x^k[n_x+1]\hat{\mathbf{q}}_x^k[n_x-1]}{\hat{\mathbf{q}}_x^k[n_x]^2}\in\mathbb{C}^{N_x-2},
\end{align}
for $n_x\in \{-\frac{N_x-1}{2}+1,\cdots,\frac{N_x-1}{2}-1\}$ with $x\in\{\H,\V\}$, $\gamma_k$ can be estimated by solving the following least squares (LS) problem
\begin{equation}
    \hat{\gamma}_k = \arg\min_{\gamma} \sum_{x\in\{\H,\V\}} \|\brho_x^k - e^{-j2\gamma d_x^2} \mathbf{1}_{N_x-2}\|_2^2.
\end{equation}
Finally, the range estimate is given by $\hat{r}_k = \frac{\pi}{\lambda \hat{\gamma}_k}$.

\noindent\textbf{3) Path gain estimation}: 
With $\{\hat{\psi}_k,\hat{\theta}_k,\hat{r}_k\}_{k=1}^K$, the RIS response is recovered as $\hat{\mathbf{b}}_k = \mathbf{b}(\hat{\psi}_k,\hat{\theta}_k,\hat{r}_k)$.
Inserting $\{\hat{\mathbf{b}}_k\}_{k=1}^K$ into (\ref{eqn:y_proj}),
the effective gains $[\tilde{\eta}_k,\cdots,\tilde{\eta}_K]^T$ are then obtained from least squares
\begin{equation}
    [\hat{\tilde{\eta}}_1,\cdots,\hat{\tilde{\eta}}_K]^T= (\hat{\bS}^H\,\hat{\bS})^{-1}\hat{\bS}^H\mathbf{y},
\end{equation}
where $\hat{\bS} =  \bar{\mathbf{H}}[\hat{\mathbf{b}}_1, \cdots, \hat{\mathbf{b}}_K]\in\mathbb{C}^{ML\times K}$.
If the power $p_k$ is known, then the path gain is estimated by $\hat{\eta}_k = \hat{\tilde{\eta}}_k/\sqrt{p_k}$.

\section{Simulation Results}
\label{sec:sim}
We conduct numerical simulations to evaluate the proposed super-resolution method on a RIS-assisted ISAC setup with $N_\H=N_\V=15$, $M=15$, and the carrier frequency $f_c=1\,\mathrm{GHz}$, i.e., $\lambda=0.3\,\mathrm{m}$.
The inter-element spacings are $d_\H=d_\V = \lambda/2$.
The RIS-BS link follows a free-space path-loss model with a separation of $6\,\mathrm{m}$.
Each UE transmits $L=10$ pilot symbols. 
In the proposed method, subspace matrices $\bB_\H$ and $\bB_\V$ are constructed by discretizing the quadratic phase with $J_\H = J_V = 3$.
We compare our proposed algorithm against three methods: (i) the OMP algorithm with a polar-domain dictionary \cite{Cui-ToC22}, (ii) the 3D-MUSIC algorithm \cite{Ramezani-GLOBECOM24}, (iii) an idealized 3D-MUSIC that assumes direct access to RIS-side data (``3D-MUSIC(RIS)''), which serves as a lower bound for the subspace methods but is not physically realizable due to the RIS’s reflection-only operation.
All the results are averaged over 50 trials.

Fig. \ref{fig:RMSEvsSNR} shows the root mean square errors (RMSEs) of the azimuth-elevation-range parameters with respect to varying SNRs. Here $K=2$ UEs are randomly distributed with $\psi\in [-\pi/3,\pi/3]$, $\theta\in [-\pi/6,\pi/6]$, and $r\in [3\rm{m},15\rm{m}]$, ensuring operation in the radiating near field.
The results show that, as SNR increases, the RMSE of the proposed method decreases significantly, whereas those of the other three methods remain nearly flat for SNRs exceeding $0\,\mathrm{dB}$.
Specifically, for azimuth and elevation estimation, the proposed method outperforms OMP and 3D-MUSIC when the SNR is greater than $-5\,\mathrm{dB}$ and surpasses 3D-MUSIC(RIS) when the SNR exceeds $0\,\mathrm{dB}$.
For range estimation, our method consistently attains the lowest RMSE across all SNRs.
Fig. \ref{fig:RMSEvsK} demonstrates the average positioning error $\sqrt{1/K\sum_{k=1}^K\|\mathbf{p}_k - \hat{\mathbf{p}}_k\|^2_2}$ versus the number of UEs at a fixed SNR of $15\,\mathrm{dB}$, where $\hat{\mathbf{p}}_k = \hat{r}_k(\cos(\hat{\psi}_k)\cos(\hat{\theta}_k),\sin(\hat{\psi}_k)\cos(\hat{\theta}_k),\sin(\hat{\theta}_k))$ is the estimated Cartesian coordinate of the $k$th UE.
The proposed method consistently outperforms the other methods as $K$ increases, indicating robustness in multi-user scenarios with densely located UEs.

\section{Conclusions}
This work presents a high-resolution parameter estimation method for RIS-assisted localization in the radiating NF region.
By exploiting the Fresnel approximation, the RIS response is equivalently represented as the element-wise product of a 2D far-field steering vector and a range-dependent quadratic-phase chirp, enabling a super-resolution formulation that jointly estimates azimuth, elevation, and range.
Simulations show that the proposed method achieves superior parameter estimation performance than those of subspace and CS methods.
Future work includes developing faster solvers tailored to our problem and optimizing RIS phase patterns for enhanced accuracy and robustness.

\vfill\pagebreak




\bibliographystyle{IEEEtran}
\bibliography{IEEEabrv,strings}

\begin{thebibliography}{10}
\providecommand{\url}[1]{#1}
\csname url@samestyle\endcsname
\providecommand{\newblock}{\relax}
\providecommand{\bibinfo}[2]{#2}
\providecommand{\BIBentrySTDinterwordspacing}{\spaceskip=0pt\relax}
\providecommand{\BIBentryALTinterwordstretchfactor}{4}
\providecommand{\BIBentryALTinterwordspacing}{\spaceskip=\fontdimen2\font plus
\BIBentryALTinterwordstretchfactor\fontdimen3\font minus \fontdimen4\font\relax}
\providecommand{\BIBforeignlanguage}[2]{{%
\expandafter\ifx\csname l@#1\endcsname\relax
\typeout{** WARNING: IEEEtran.bst: No hyphenation pattern has been}%
\typeout{** loaded for the language `#1'. Using the pattern for}%
\typeout{** the default language instead.}%
\else
\language=\csname l@#1\endcsname
\fi
#2}}
\providecommand{\BIBdecl}{\relax}
\BIBdecl

\bibitem{Renzo-JSAC20}
M.~Di~Renzo, A.~Zappone, M.~Debbah, M.-S. Alouini, C.~Yuen, J.~de~Rosny, and S.~Tretyakov, ``Smart radio environments empowered by reconfigurable intelligent surfaces: How it works, state of research, and the road ahead,'' \emph{IEEE Journal on Selected Areas in Communications}, vol.~38, no.~11, pp. 2450--2525, 2020.

\bibitem{Emil-SPM22}
E.~Björnson, H.~Wymeersch, B.~Matthiesen, P.~Popovski, L.~Sanguinetti, and E.~de~Carvalho, ``Reconfigurable intelligent surfaces: A signal processing perspective with wireless applications,'' \emph{IEEE Signal Processing Magazine}, vol.~39, no.~2, pp. 135--158, 2022.

\bibitem{Wymeersch-VTM20}
H.~Wymeersch, J.~He, B.~Denis, A.~Clemente, and M.~Juntti, ``Radio localization and mapping with reconfigurable intelligent surfaces: Challenges, opportunities, and research directions,'' \emph{IEEE Vehicular Technology Magazine}, vol.~15, no.~4, pp. 52--61, 2020.

\bibitem{Pan-JSTSP23}
Y.~Pan, C.~Pan, S.~Jin, and J.~Wang, ``{RIS}-aided near-field localization and channel estimation for the terahertz system,'' \emph{IEEE Journal of Selected Topics in Signal Processing}, vol.~17, no.~4, pp. 878--892, 2023.

\bibitem{Yan-TGCN25}
H.~Yan, H.~Chen, W.~Liu, S.~Yang, G.~Wang, and C.~Yuen, ``{RIS}-enabled joint near-field {3D} localization and synchronization in siso multipath environments,'' \emph{IEEE Transactions on Green Communications and Networking}, vol.~9, no.~1, pp. 367--379, 2025.

\bibitem{Liu-JSAC22}
F.~Liu, Y.~Cui, C.~Masouros, J.~Xu, T.~X. Han, Y.~C. Eldar, and S.~Buzzi, ``Integrated sensing and communications: Toward dual-functional wireless networks for {6G} and beyond,'' \emph{IEEE Journal on Selected Areas in Communications}, vol.~40, no.~6, pp. 1728--1767, 2022.

\bibitem{stoica2005spectral}
P.~Stoica and R.~Moses, \emph{Spectral Analysis of Signals}.\hskip 1em plus 0.5em minus 0.4em\relax Pearson Prentice Hall, 2005.

\bibitem{Malioutov-TSP05}
D.~Malioutov, M.~Cetin, and A.~Willsky, ``A sparse signal reconstruction perspective for source localization with sensor arrays,'' \emph{IEEE Transactions on Signal Processing}, vol.~53, no.~8, pp. 3010--3022, 2005.

\bibitem{Ramezani-GLOBECOM24}
P.~Ramezani, A.~Kosasih, and E.~Björnson, ``An efficient modified {MUSIC} algorithm for {RIS}-assisted near-field localization,'' in \emph{GLOBECOM 2024 - 2024 IEEE Global Communications Conference}, 2024, pp. 4430--4435.

\bibitem{Cui-ToC22}
M.~Cui and L.~Dai, ``Channel estimation for extremely large-scale {MIMO}: Far-field or near-field?'' \emph{IEEE Transactions on Communications}, vol.~70, no.~4, pp. 2663--2677, 2022.

\bibitem{Wei-CCom22}
X.~Wei, L.~Dai, Y.~Zhao, G.~Yu, and X.~Duan, ``Codebook design and beam training for extremely large-scale {RIS}: Far-field or near-field?'' \emph{China Communications}, vol.~19, no.~6, pp. 193--204, 2022.

\bibitem{Liu-ICC04}
S.~Liu, X.~Yu, Z.~Gao, and D.~W.~K. Ng, ``{DPSS}-based codebook design for near-field {XL-MIMO} channel estimation,'' in \emph{ICC 2024 - IEEE International Conference on Communications}, 2024, pp. 3864--3870.

\bibitem{Yang-TIT16}
D.~Yang, G.~Tang, and M.~B. Wakin, ``Super-resolution of complex exponentials from modulations with unknown waveforms,'' \emph{IEEE Transactions on Information Theory}, vol.~62, no.~10, pp. 5809--5830, 2016.

\bibitem{Chi-TSP2015}
Y.~Chi and Y.~Chen, ``Compressive two-dimensional harmonic retrieval via atomic norm minimization,'' \emph{IEEE Transactions on Signal Processing}, vol.~63, no.~4, pp. 1030--1042, 2015.

\bibitem{Suliman-TSP22}
M.~A. Suliman and W.~Dai, ``Blind two-dimensional super-resolution and its performance guarantee,'' \emph{IEEE Transactions on Signal Processing}, vol.~70, pp. 2844--2858, 2022.

\bibitem{Li-ICASSP25}
J.~Li, Q.~Chen, and F.~Xi, ``A blind super-resolution method for near-field channel estimation with angle-range recovery,'' in \emph{ICASSP 2025 - 2025 IEEE International Conference on Acoustics, Speech and Signal Processing (ICASSP)}, 2025, pp. 1--5.

\bibitem{Slepian-1978}
D.~Slepian, ``Prolate spheroidal wave functions, fourier analysis, and uncertainty-v: the discrete case,'' \emph{The Bell System Technical Journal}, vol.~57, no.~5, pp. 1371--1430, 1978.

\bibitem{Zhu-2017}
Z.~Zhu and M.~B. Wakin, ``Approximating sampled sinusoids and multiband signals using multiband modulated dpss dictionaries,'' \emph{Journal of Fourier Analysis and Applications}, vol.~23, p. 1263–1310, Dec. 2017.

\bibitem{cvx}
M.~Grant and S.~Boyd, ``{CVX}: Matlab software for disciplined convex programming, version 2.1,'' \url{https://cvxr.com/cvx}, Mar. 2014.

\bibitem{Ran-ICASSP21}
Y.~Ran and W.~Dai, ``Fast and robust {ADMM} for blind super-resolution,'' in \emph{ICASSP 2021 - 2021 IEEE International Conference on Acoustics, Speech and Signal Processing (ICASSP)}, 2021, pp. 5150--5154.

\bibitem{Li-TSP24}
R.~Li and D.~Cabric, ``A coordinate descent approach to atomic norm denoising,'' \emph{IEEE Transactions on Signal Processing}, vol.~72, pp. 5077--5090, 2024.

\bibitem{Yang-TIT2016}
Z.~Yang, L.~Xie, and P.~Stoica, ``Vandermonde decomposition of multilevel {Toeplitz} matrices with application to multidimensional super-resolution,'' \emph{IEEE Transactions on Information Theory}, vol.~62, no.~6, pp. 3685--3701, 2016.

\end{thebibliography}

\end{document}